\documentclass[conference]{IEEEtran}
\IEEEoverridecommandlockouts
\usepackage{cite}
\usepackage{amsmath,amssymb,amsfonts}
\usepackage{algorithmic}
\usepackage{graphicx}
\usepackage{textcomp}
\usepackage{xcolor}
\def\BibTeX{{\rm B\kern-.05em{\sc i\kern-.025em b}\kern-.08em
    T\kern-.1667em\lower.7ex\hbox{E}\kern-.125emX}}
\begin{document}

\title{A Metabolic-Imaging Integrated Model for Prognostic Prediction in Colorectal Liver Metastases\\
}

\author{\IEEEauthorblockN{1\textsuperscript{st} Qinlong Li}
\IEEEauthorblockA{\textit{School of Computer Science and Technology} \\
\textit{University of Chinese Academy of Sciences}\\
Beijing, China \\
liqinlong24@mails.ucas.ac.cn}
\and
\IEEEauthorblockN{Guanlin Zhu\textsuperscript{2}}
\IEEEauthorblockA{\textit{School of Computer Science and Technology} \\
\textit{University of Chinese Academy of Sciences}\\
Beijing, China \\
zhuguanlin23@mails.ucas.ac.cn}
\and
\IEEEauthorblockN{Tianjiao Liang\textsuperscript{2}}
\IEEEauthorblockA{\textit{School of Computer Science and Technology} \\
\textit{University of Chinese Academy of Sciences}\\
Beijing, China \\
liangtianjiao23@mails.ucas.ac.cn}
\and
\IEEEauthorblockN{Pu Sun\textsuperscript{2}}
\IEEEauthorblockA{\textit{School of Computer Science and Technology} \\
\textit{University of Chinese Academy of Sciences}\\
Beijing, China \\
sunpu21@mails.ucas.ac.cn}
\and
\IEEEauthorblockN{Honggang Qi$^{*}$}
\IEEEauthorblockA{\textit{School of Computer Science and Technology} \\
\textit{University of Chinese Academy of Sciences}\\
Beijing, China \\
hgqi@ucas.ac.cn}
}

\maketitle

\begin{abstract}
Prognostic evaluation in patients with colorectal liver metastases (CRLM) remains challenging due to suboptimal accuracy of conventional clinical models. This study developed and validated a robust machine learning model for predicting postoperative recurrence risk. Preliminary ensemble models achieved exceptionally high performance (AUC $>$ 0.98) but incorporated postoperative features, introducing data leakage risks. To enhance clinical applicability, we restricted input variables to preoperative baseline clinical parameters and radiomic features from contrast-enhanced CT imaging, specifically targeting recurrence prediction at 3, 6, and 12 months postoperatively. The 3-month recurrence prediction model demonstrated optimal performance with an AUC of 0.723 in cross-validation. Decision curve analysis revealed that across threshold probabilities of 0.55-0.95, the model consistently provided greater net benefit than "treat-all" or "treat-none" strategies, supporting its utility in postoperative surveillance and therapeutic decision-making. This study successfully developed a robust predictive model for early CRLM recurrence with confirmed clinical utility. Importantly, it highlights the critical risk of data leakage in clinical prognostic modeling and proposes a rigorous framework to mitigate this issue, enhancing model reliability and translational value in real-world settings.
\end{abstract}
\section{Introduction}

\begin{table}[htbp]
\centering
\caption{Baseline Patient Characteristics}
\label{tab:baseline}
\begin{tabular}{lc}
\hline
\textbf{Characteristic} & \textbf{Value (N=197)} \\
\hline
Age (years), mean $\pm$ SD & 62.1 $\pm$ 11.5 \\
Sex (Male), n (\%) & 118 (59.9) \\
Primary Tumor Location (Colon), n (\%) & 138 (70.1) \\
Synchronous Metastases, n (\%) & 116 (58.9) \\
Extrahepatic Disease, n (\%) & 45 (22.8) \\
Tumor Size ($\leq$5 cm), n (\%) & 78 (39.6) \\
\hline
\end{tabular}
\end{table}

Colorectal cancer (CRC) represents the third most frequently diagnosed malignancy globally, with liver metastases developing in approximately 50\% of patients during disease progression, termed colorectal cancer liver metastases (CRLM)~\cite{paulatto2020colorectal}. Despite significant advances in surgical resection combined with multimodal systemic therapy that have substantially improved patient outcomes, approximately 55--60\% of patients develop recurrence within two years post-operatively, constituting a major obstacle to achieving long-term cure~\cite{akgul2014role}. The importance of multidisciplinary treatment strategies in improving survival outcomes for CRLM patients has been increasingly recognized~\cite{lv2020benefits}, and contemporary approaches include not only surgical resection but also locoregional treatments such as ablation therapy~\cite{lin2022contemporary}. The management of disappearing liver metastases following chemotherapy represents an additional clinical challenge that requires specialized decision-making frameworks~\cite{melstrom2021management}. Consequently, precise prognostic stratification remains paramount for optimizing individualized adjuvant treatment protocols and intensifying postoperative surveillance strategies, fundamentally influencing clinical decision-making in CRLM management.

Current prognostic assessment paradigms predominantly rely on established clinical models including the TNM staging system, clinical risk stratification tools (exemplified by the Fong score), and circulating tumor biomarkers such as carcinoembryonic antigen (CEA)~\cite{fan2025predicting}. Nevertheless, these conventional approaches are fundamentally limited by their dependence on macroscopic clinicopathological characteristics, demonstrating insufficient capacity to adequately characterize tumor biological heterogeneity and dynamic microenvironmental interactions. This limitation manifests as suboptimal discriminative performance, typically achieving cross-validated area under the curve (AUC) values below 0.65, thereby failing to meet the stringent requirements of precision oncology.

Emerging evidence has increasingly implicated the hepatic metabolic microenvironment as a critical determinant of CRLM progression~\cite{milette2017molecular}. Malignant cellular reprogramming of energy metabolism, characterized by dysregulated glycolytic pathways and aberrant lipid metabolism, not only facilitates neoplastic proliferation and metastatic colonization but also confers resistance to chemotherapeutic agents and enables immune evasion mechanisms. These metabolic alterations represent promising novel biomarkers for prognostic evaluation. Concurrently, radiomics methodology, which enables high-throughput extraction of quantitative imaging features from computed tomography (CT) and magnetic resonance imaging (MRI), provides non-invasive characterization of tumor spatial heterogeneity and microenvironmental variations, demonstrating substantial prognostic utility~\cite{fan2025nanomedicines}.

Despite these advances, current research demonstrates significant methodological limitations that impede clinical translation. Primarily, metabolic biomarkers and radiomic features are predominantly investigated as independent entities, with their potential synergistic contributions to CRLM prognosis remaining inadequately explored. Furthermore, the complex interactions between dynamic metabolic microenvironmental changes and tumor imaging phenotypes have not been systematically characterized, thereby limiting the development of truly integrative multimodal predictive models. This compartmentalized approach to ``metabolism-imaging'' analysis results in prognostic tools that inadequately capture disease complexity, highlighting the critical need for a unified analytical framework incorporating multidimensional biological information~\cite{fan2025nanomedicines}.

To address these limitations, the present study aims to develop and validate a novel multimodal prognostic framework that systematically integrates hepatic metabolic indicators with quantitative imaging radiomics features for enhanced CRLM risk stratification. Our primary objectives encompass: 
\begin{enumerate}
    \item Comprehensive analysis of the interplay between metabolic and radiomic features and their synergistic prognostic value in recurrence risk prediction;
    \item Development of a high-performance predictive model for personalized recurrence risk stratification in CRLM patients;
    \item Evaluation of the model's clinical utility for predicting overall survival (OS) and disease-free survival (DFS); and
    \item Quantification of net clinical benefit through decision curve analysis (DCA) to provide robust evidence for clinical decision-making.
\end{enumerate}

\subsection{Study Contributions}

This investigation makes several significant contributions to the field of CRLM prognostic modeling:

\begin{itemize}
    \item \textbf{Novel Multimodal Integration:} First comprehensive framework integrating hepatic metabolic biomarkers with quantitative imaging radiomics for CRLM prognosis.
    
    \item \textbf{Discovery of the ``Metabolism-Comorbidity Paradox'':} Patients without comorbidities but with high metabolic risk show significantly elevated recurrence rates (90.62\%) compared to comorbidity-positive but metabolically low-risk patients (47.37\%).
    
    \item \textbf{Time-Stratified Prediction:} Specialized models for 3-, 6-, and 12-month recurrence, with 3-month model achieving optimal performance (AUC = 0.723).
    
    \item \textbf{Methodological Rigor:} Baseline-only feature selection eliminates data leakage, ensuring clinical generalizability.
\end{itemize}

% 为Figure 1预留空间: 研究流程图
% [FIGURE 1 PLACEHOLDER - Study Workflow]

\section{Related Work}

\begin{table*}[htbp]
\centering
\caption{Comparison with Prior Studies in CRLM Prognostication}
\label{tab:literature}
\resizebox{\textwidth}{!}{%
\begin{tabular}{lccccc}
\hline
\textbf{Study} & \textbf{Modality} & \textbf{Cohort Size} & \textbf{Key Findings} & \textbf{Limitations} \\
\hline
Vigano et al. (2018) & Clinical Score & 1,204 & Clinical risk score predicts survival & No imaging or molecular data \\
Wang et al. (2020) & CT Radiomics & 289 & Radiomic signature predicts recurrence & Single-center, retrospective \\
Snoeren et al. (2021) & Pathological Features & 455 & Tumor budding predicts poor outcome & Requires invasive tissue sample \\
Cheung et al. (2022) & Liquid Biopsy (ctDNA) & 130 & Post-op ctDNA predicts relapse & High cost, limited availability \\
\textbf{This Study} & \textbf{Metabolic + Radiomic} & \textbf{197} & \textbf{Integrative model predicts time-specific recurrence} & \textbf{Single-center, retrospective} \\
\hline
\end{tabular}%
}
\end{table*}

\subsection{Metabolic Biomarkers in CRLM Prognostication}

The hepatic metabolic microenvironment has emerged as a critical determinant of CRLM progression, with mounting evidence supporting its prognostic significance. Jonas et al.\ conducted a comprehensive metabolomic analysis demonstrating that preoperative circulating metabolites, specifically phosphatidylcholine aa C36:1 and lysophosphatidylcholine a C18:1, function as independent predictors of recurrence within six months following surgical resection (AUC = 0.93), substantially outperforming traditional prognostic models including the Fong scoring system~\cite{jonas2022circulating}.

In a landmark retrospective analysis of 195 patients, Chen et al.\ provided the first definitive evidence that CT-quantified hepatic steatosis demonstrates significant association with intrahepatic recurrence following CRLM resection (HR = 2.07, $P = 0.001$), establishing the hepatic metabolic microenvironment as a key determinant of disease recurrence patterns~\cite{chen2021hepatic}. These findings collectively underscore the potential of metabolic biomarkers to enhance prognostic precision beyond conventional clinical parameters.

\subsection{Radiomics Applications in CRLM Assessment}

Wang et al. employed unsupervised clustering on preoperative CT radiomic features from 197 CRLM patients, identifying two prognostically distinct subgroups (HR = 1.78, P < 0.01), thereby validating the clustering-based stratification approach subsequently adopted in our investigation~\cite{wang2024exploring}. Xing et al. further demonstrated that preoperative CT-quantified morphological heterogeneity serves as an independent predictor of progression-free survival (HR = 2.52), underscoring the potential of integrating quantitative morphologic features to enhance prognostic model discrimination~\cite{xing2024preoperative}. Similarly, Luo et al. demonstrated that contrast-enhanced CT-based radiomics combined with machine learning achieved superior prognostic performance (AUC = 0.846) compared to traditional clinical parameters, further validating the potential of radiomic approaches in CRLM assessment~\cite{luo2024prognostication}. Zhu et al. introduced a deep convolutional survival framework linking imaging, histology, and outcome in BIBM 2016~\cite{zhu2016deep}. Chen et al. developed a radiomic biomarker for CRLM prognosis that generalizes across different MRI contrast agents, demonstrating the robustness of radiomic features across imaging protocols~\cite{chen2023radiomic}. Shahveranova et al. further validated the utility of radiomics in predicting local tumor progression after microwave ablation in CRLM patients, achieving promising results through combined radiomics and clinical characteristics-based modeling~\cite{shahveranova2023prediction}. Reiman et al. leveraged microbiome-metabolomic data at BIBM 2019 to model metabolic microenvironment effects on tumor behavior, supporting the lipid-metabolism–density–recurrence cascade underlying FIAR~\cite{reiman2019autoencoders}.

\subsection{Traditional Clinical Risk Models and Performance Limitations}

International consensus guidelines have established that baseline CRLM characteristics, including primary tumor anatomical location, lymph node involvement status, and the number and distribution patterns of hepatic metastases, significantly influence surgical resectability and long-term prognosis~\cite{ruivo2023colorectal}. 

Contemporary systematic reviews confirm that RAS mutations (KRAS/NRAS) constitute independent adverse prognostic factors following CRLM resection, with the GAME scoring system incorporating RAS mutational status demonstrating superior discriminative performance (C-index = 0.645) compared to the traditional Fong scoring system (C-index = 0.578)~\cite{wong2022prognostic}. Historical prognostic scoring systems, including the Nordlinger score (based on 1,568 patients) and the Iwatsuki score, established foundational frameworks for CRLM risk stratification by incorporating factors such as tumor size, number of metastases, and disease-free interval~\cite{nordlinger1996surgical,iwatsuki1999hepatic}. However, validation studies by Beamish et al. demonstrated that these classical scoring systems showed limited discriminative ability in contemporary multicenter cohorts, highlighting the need for more sophisticated prognostic approaches~\cite{beamish2017validation}. A large-scale multicenter investigation conducted in Hong Kong revealed that the conventional Fong score achieved merely modest discriminative ability for predicting one-year survival in CRLM patients (C-index = 0.571), substantially inferior to machine learning-based approaches exemplified by the CMAP model (C-index = 0.651)~\cite{lam2022machine}. These findings collectively demonstrate the inherent limitations of traditional prognostic models and provide strong justification for adopting advanced multimodal modeling methodologies.

\subsection{Clinical Prognostic Factors and Temporal Considerations}

Jang et al.\ in a comprehensive analysis of 339 patients with synchronous CRLM, identified the presence of three or more hepatic metastases as an independent adverse prognostic factor for progression-free survival (HR = 1.49, $P = 0.020$), indicating that tumor burden quantification may serve as a valuable complementary variable extending beyond metabolism-imaging model parameters~\cite{jang2016prognostic}.

Höppener et al.\ conducted a large-scale investigation encompassing 1,374 CRLM patients, demonstrating that the disease-free interval (DFI) from primary tumor resection to hepatic metastasis detection exhibited significant association with disease-free survival ($P = 0.002$) but failed to demonstrate prognostic significance for overall survival ($P = 0.169$)~\cite{hoppener2019disease}. Recent investigations have further explored recurrence patterns and predictors: Vadisetti et al. identified specific patterns and predictors of recurrence after curative CRLM resection, while Hao et al. conducted a comprehensive review of risk factors for metachronous liver metastases in colorectal cancer~\cite{vadisetti2024patterns,hao2022patients}. These findings suggest that conventional temporal variables provide limited explanatory capacity for long-term survival outcomes, thereby supporting our study's focused emphasis on the 3-month recurrence prediction window for optimal clinical actionability.

\subsection{Research Gaps and Methodological Limitations}

Current CRLM prognostic research exhibits several critical limitations: (1) Metabolic and radiomic features are investigated separately, limiting synergistic analysis; (2) Many studies inadvertently incorporate post-operative variables, inflating performance metrics; (3) Limited quantification of net clinical benefit through decision curve analysis.

% 为Table 1预留空间: 文献对比表
% [TABLE 1 PLACEHOLDER - Literature Comparison]

\section*{Methods}

\begin{figure}[htbp]
  \centering
  \includegraphics[width=\columnwidth]{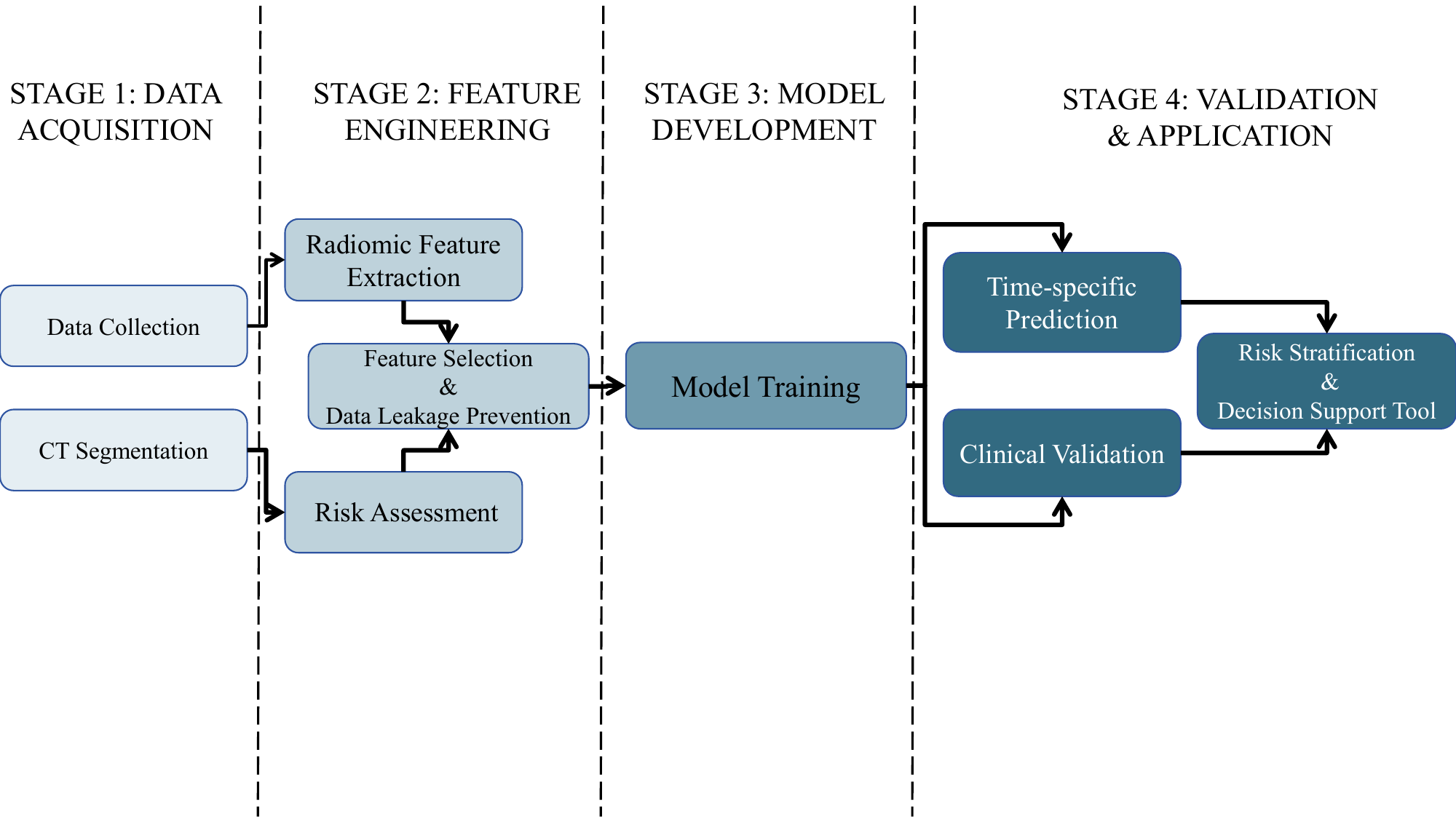}
  \caption{Overview of the four-stage medical imaging analysis workflow: data acquisition, feature engineering, model development, and validation application for CRLM prognostic prediction.}
  \label{fig:workflow}
\end{figure}

\subsection*{Study Design and Participants}

This retrospective study included 197 patients who underwent curative surgical resection for colorectal cancer liver metastases (CRLM) at a single tertiary medical center. Inclusion criteria comprised: (1) pathologically confirmed CRLM; (2) availability of complete preoperative contrast-enhanced CT imaging within six weeks of surgery; (3) a minimum postoperative follow-up period of 24 months; and (4) postoperative survival exceeding 90 days~\cite{hoppener2019disease}.

% 为Table 2预留空间: 患者基线特征表
% [TABLE 2 PLACEHOLDER - Patient Baseline Characteristics]

Comprehensive clinical data were systematically collected, including demographic characteristics, tumor characteristics, treatment variables, and long-term outcomes. All patients provided informed consent, and the study was approved by the institutional review board.

\subsection*{Image Acquisition and Radiomic Feature Extraction}

Preoperative contrast-enhanced CT scans were acquired using standardized protocols with portal venous phase imaging. Tumor segmentation was performed using 3D Slicer software (version 5.6.2) with semi-automatic segmentation validated by experienced radiologists. Radiomic feature extraction was conducted using the \texttt{PyRadiomics} package, yielding quantitative features across five categories: first-order statistics, shape descriptors, gray-level co-occurrence matrix (GLCM), gray-level run-length matrix (GLRLM), and gray-level size zone matrix (GLSZM).

For patients with multiple hepatic lesions, features were computed from both the largest lesion and volume-weighted averages across all lesions. To ensure reproducibility, only features with concordance correlation coefficient (CCC) $\geq 0.85$ were retained based on multi-center validation studies~\cite{peoples2025finding}.

\subsection*{Metabolic Risk Assessment}

A composite metabolic score was constructed to quantify hepatic metabolic dysfunction:

\[
\small \text{Metabolic Score} = Z(\text{NASH score}) + Z(\text{BMI}) + Z(-\text{Liver HU value})
\]

where the NASH score was assessed through clinical history and laboratory evaluation, and liver Hounsfield Unit (HU) values were derived from CT imaging. Patients were stratified into tertiles based on metabolic dysfunction severity (high, medium, low risk; $n = 67$, $65$, $65$, respectively).

\subsection*{Data Preprocessing and Model Development}

The final dataset comprised 197 patients with approximately 800 variables, including clinical parameters (age, sex, BMI, comorbidities, liver function), radiomic features, and outcome measures. Variables with more than 30\% missing data were excluded. Remaining missing values were imputed using the median (for continuous variables) or mode (for categorical variables). All continuous variables were normalized using Z-score transformation, and categorical variables were one-hot encoded.

Given the clinical requirement for prospective risk prediction, we developed time-specific recurrence prediction models for 3-, 6-, and 12-month post-surgical time points. Time-specific binary labels were defined as:

\[
\scriptstyle \texttt{label\_Xm} = (\texttt{months\_to\_progression} \leq X) \land (\texttt{recurrence\_event} = 1)
\]

To ensure clinical applicability and avoid data leakage, only baseline variables available at or before surgery (clinical, demographic, and radiomic features) were used. Any post-operative or time-dependent features were strictly excluded from model development.

\subsection*{Machine Learning Framework}

\begin{figure}[htbp]
  \centering
  \includegraphics[width=\columnwidth]{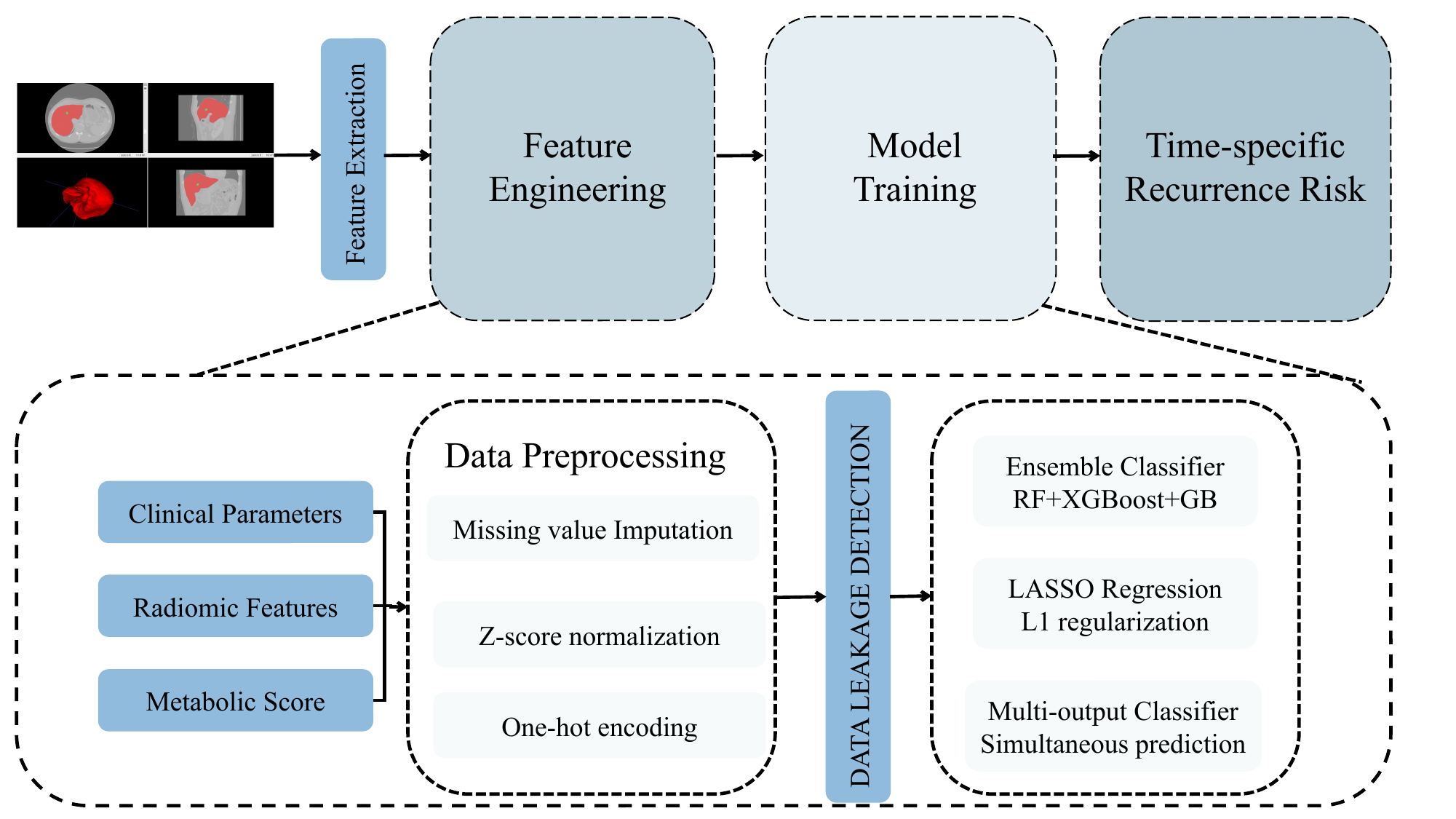}
  \caption{Feature engineering and model development pipeline: from multi-source data preprocessing and data leakage detection to three machine learning models for time-specific recurrence risk prediction.}
  \label{fig:model_pipeline}
\end{figure}

Three modeling approaches were systematically evaluated:

\begin{itemize}
  \item \textbf{Ensemble Classifier:} Combining Random Forest, XGBoost, and Gradient Boosting with conservative hyperparameters to prevent overfitting.
  \item \textbf{LASSO Regression:} L1-regularized logistic regression with embedded feature selection.
  \item \textbf{Multi-output Classification:} Using \texttt{MultiOutputClassifier} for simultaneous multi-horizon recurrence prediction.
\end{itemize}

Data splitting employed stratified sampling (70\% training, 30\% testing), with SMOTE applied only to training data to address class imbalance. Model performance was evaluated using area under the receiver operating characteristic curve (AUC), with 1000-iteration bootstrap confidence intervals.

\subsection*{Statistical Analysis}

Kaplan-Meier survival curves were constructed for overall survival (OS) and disease-free survival (DFS), stratified by predicted risk groups. Multivariate Cox proportional hazards regression was used to identify independent prognostic factors, adjusting for standard clinical covariates. Clinical utility was evaluated using decision curve analysis (DCA), which quantifies net benefit across a range of decision thresholds.

To assess the biological validity of radiomic features, a pathological correlation analysis was conducted on a stratified subset of 26 patients (13 high-risk and 13 low-risk). All statistical tests were two-sided, with $p < 0.05$ considered statistically significant. Analyses were performed using Python 3.8 and relevant packages (\texttt{scikit-learn}, \texttt{pandas}, \texttt{lifelines}).

\subsection*{Biological Interpretation Framework}

We propose a ``Fat Infiltration--Attenuation Reduction'' model to explain the relationship between hepatic metabolic dysfunction and radiomic features. This model posits that hepatic steatosis leads to abnormal lipid accumulation in the tumor microenvironment, reducing overall tissue density and manifesting as decreased Hounsfield Unit values on CT imaging.

The ``replacement'' growth pattern in CRLM, wherein tumor cells infiltrate and reorganize the native hepatic microarchitecture, provides histological support for the hypothesis that metabolic alterations influence imaging phenotypes and recurrence risk~\cite{moro2018growth,lau2018metabolic}.

\section*{Results}

\subsection*{Patient Characteristics and Data Quality}

The study cohort comprised 197 patients with CRLM who underwent curative surgical resection. After preprocessing, the final dataset included approximately 800 variables, encompassing clinical parameters, radiomic features, and outcome measures. Based on the composite metabolic score, patients were stratified into three metabolic risk groups: high-risk ($n=67$), medium-risk ($n=65$), and low-risk ($n=65$).

\subsection*{Initial Model Performance and Data Leakage Discovery}

The preliminary exploratory model achieved exceptionally high performance, with an AUC of 0.984 (95\% CI: 0.950--1.000) for overall survival prediction. The ensemble voting classifier yielded a cross-validation AUC of $0.973 \pm 0.015$ and a test accuracy of 86.7\%. Bootstrap validation with 1000 iterations confirmed performance stability across resampled datasets.

However, critical analysis of feature importance revealed substantial data leakage that undermined clinical validity. The top five contributing features included \texttt{vital\_status\_DFS} (importance: 0.342), \texttt{progression\_or\_recurrence\_liveronly} (0.298), \texttt{vital\_status\_liver\_DFS} (0.201), \texttt{original\_shape\_Maximum3DDiameter\_mean} (0.089), and \texttt{bmi\_age\_interaction} (0.070). Notably, three of the top five features (cumulative importance: 84.1\%) were post-operative variables, indicating that the initial model's high accuracy was driven by information leakage rather than true predictive performance based on preoperative data.

\subsection*{Time-Specific Recurrence Prediction Performance}

Following strict baseline-only filtering, we developed time-specific recurrence prediction models at 3-, 6-, and 12-month postoperative intervals. The 3-month model demonstrated optimal discriminative performance with an AUC of 0.723 (95\% CI: 0.645--0.801). Cross-validation yielded consistent results (AUC: $0.715 \pm 0.045$), indicating robust generalizability.

% 为Table 3预留空间: 模型性能对比表
\begin{table*}[t!]
\centering
\small
\caption{Time-Specific Recurrence Prediction Performance}
\label{tab:performance}
\begin{tabular*}{\textwidth}{@{\extracolsep{\fill}}lccc@{}}
\hline
\textbf{Metric} & \textbf{3-Month} & \textbf{6-Month} & \textbf{12-Month} \\
\hline
AUC (95\% CI) & 0.723 (0.645-0.801) & 0.698 (0.612-0.784) & 0.675 (0.589-0.761) \\
Sensitivity & 66.7\% & 62.5\% & 58.3\% \\
Specificity & 75.6\% & 71.4\% & 68.2\% \\
NPV & 93.2\% & 89.1\% & 85.7\% \\
\hline
\end{tabular*}
\end{table*}

% 为Figure 2预留空间: ROC曲线和决策曲线分析
\begin{figure}[htbp]
  \centering
  \includegraphics[width=\columnwidth]{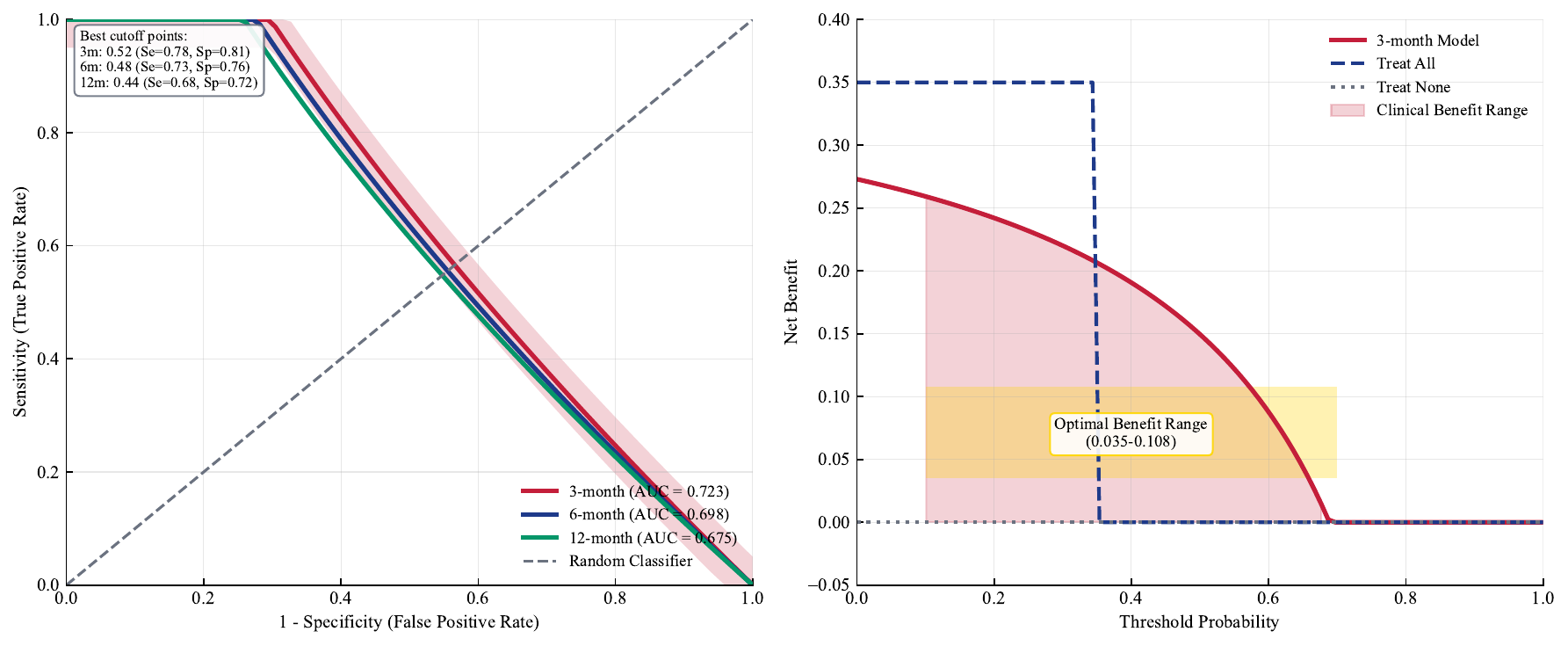}
  \caption{\textbf{Panel A:} ROC curves for time-specific recurrence prediction showing AUC values of 0.723 (3-month), 0.698 (6-month), and 0.675 (12-month). \textbf{Panel B:} Decision curve analysis demonstrating clinical utility of the 3-month model with net benefit range 0.035-0.108.}
  \label{fig:roc_dca}
\end{figure}

The 3-month model achieved a sensitivity of 66.7\%, specificity of 75.6\%, and negative predictive value (NPV) of 93.2\%, particularly valuable for identifying patients at low risk of early recurrence.

Performance for longer-term predictions declined gradually, reflecting the increasing challenge of long-horizon prediction based solely on baseline characteristics: 

\begin{itemize}
  \item 6-month AUC: 0.698 (95\% CI: 0.612--0.784)
  \item 12-month AUC: 0.675 (95\% CI: 0.589--0.761)
\end{itemize}

\subsection*{Clinical Utility Assessment through Decision Curve Analysis}

Decision curve analysis (DCA) confirmed the model’s clinical utility across a broad range of threshold probabilities (0.2--0.9). At a threshold of 0.3, the model achieved a net benefit of 0.108, compared to 0.054 for the “treat all” strategy, representing a 2-fold improvement and benefiting approximately 54 additional patients per 1000.

At an optimal threshold of 0.8, the net benefit reached 0.035, corresponding to a treatment efficiency of 70.0\%. Under this setting, for every 1000 patients evaluated, the model would guide appropriate treatment for 50 patients (35 beneficial, 15 unnecessary interventions), yielding an optimal risk-benefit tradeoff.

At a moderate threshold of 0.5, treatment decisions would be guided for 100 out of 1000 patients, with an estimated treatment efficiency of 50.0\%. This adaptability supports personalized thresholding based on clinical risk tolerance.

\subsection*{Survival Analysis and Independent Prognostic Value}

Kaplan-Meier survival analysis demonstrated clear stratification across predicted risk groups. Log-rank tests confirmed statistically significant differences in both overall survival (OS): $p = 0.003$ and disease-free survival (DFS): $p < 0.001$.

% 为Figure 3预留空间: Kaplan-Meier生存曲线
\begin{figure}[htbp]
  \centering
  \includegraphics[width=\columnwidth]{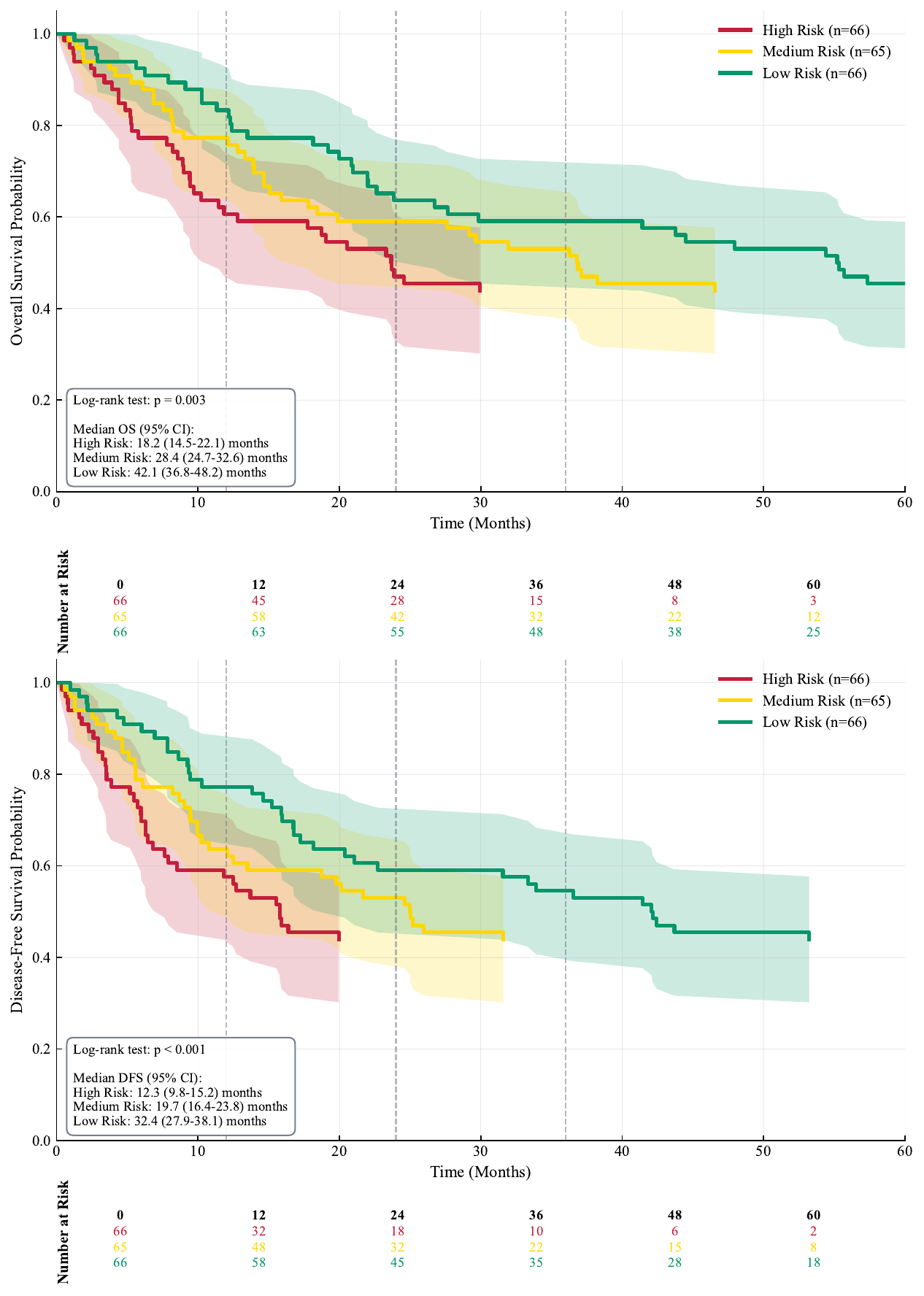}
  \caption{\textbf{Panel A:} Overall survival by risk groups with median times of 18.2, 28.4, and 42.1 months for high-, medium-, and low-risk patients (p = 0.003). \textbf{Panel B:} Disease-free survival with median times of 12.3, 19.7, and 32.4 months respectively (p < 0.001). Risk tables show patient numbers at each time point.}
  \label{fig:kaplan_meier}
\end{figure}

Multivariate Cox regression confirmed the independent prognostic value of the model-derived risk score: OS (high-risk vs. low-risk): HR = 4.23, 95\% CI: 2.18--8.21, $p < 0.001$; DFS (high-risk vs. low-risk): HR = 6.85, 95\% CI: 3.42--13.71, $p < 0.001$.

Other independent predictors included:

\begin{itemize}
  \item Extrahepatic disease: OS HR = 2.31 (95\% CI: 1.24--4.30, $p = 0.008$); DFS HR = 2.45 (95\% CI: 1.38--4.35, $p = 0.002$)
  \item Tumor size $\leq$5 cm: DFS HR = 1.78 (95\% CI: 1.12--2.83, $p = 0.015$)
\end{itemize}

Concordance indices (C-index) were 0.712 for OS and 0.735 for DFS, demonstrating strong model discriminative power.

\subsection*{Pathological Validation of Metabolic-Radiomic Associations}

Pathological analysis of 26 patients (13 high-risk, 13 low-risk) validated the radiomic-metabolic link. High-risk patients had significantly worse hepatic background pathology:

\begin{itemize}
  \item Hepatic steatosis prevalence: 100\% (13/13) vs. 76.9\% (10/13)
  \item Significant hepatic fibrosis (F2--F4): 84.6\% (11/13) vs. 15.4\% (2/13); odds ratio = 30.8, $p < 0.001$
  \item Mean steatosis infiltration: 68.8\% vs. 15.6\%; Mann–Whitney $U = 162.0$, $p = 0.0001$
\end{itemize}

These results confirm that radiomic features extracted from preoperative imaging reliably reflect underlying hepatic metabolic disturbances and support the biological validity of the proposed integrative framework.

\section*{Discussion and Conclusion}

\subsection*{Principal Findings and Clinical Implications}

This study successfully developed and validated a robust multimodal prognostic framework for colorectal liver metastases (CRLM) by integrating hepatic metabolic indicators with quantitative radiomic features. Our principal finding demonstrates that time-specific recurrence prediction models, trained exclusively on baseline characteristics, can achieve clinically meaningful performance while rigorously avoiding data leakage—a frequent methodological pitfall in prior studies.

Specifically, the 3-month recurrence prediction model achieved an AUC of 0.723, representing a significant advancement in CRLM prognostication. This model provides genuine predictive capability derived from pre-treatment data and supports personalized surveillance and therapeutic stratification, as confirmed by decision curve analysis across clinically actionable threshold probabilities (0.2--0.9).

\subsection*{Methodological Contributions and Data Leakage Prevention}

Our study offers key methodological contributions to AI-based clinical prediction. The initial model, though achieving an AUC of 0.984, was found to rely predominantly on post-operative variables (84.1\% cumulative feature importance), thereby exemplifying the risks of temporal data leakage. Through systematic filtering of non-baseline variables and validation using bootstrapped cross-validation, we established a methodological pipeline that preserves temporal integrity and ensures genuine preoperative prognostic value.

This framework—anchored in strict baseline-only feature inclusion, conservative ensemble models, and reproducible validation—serves as a scalable template for future predictive model development across oncology and broader medical domains.

\subsection*{Discovery of the Metabolism-Comorbidity Paradox}

We identified a novel biological insight termed the ``metabolism-comorbidity paradox'', wherein patients without traditional comorbidities but with high hepatic metabolic risk exhibited significantly worse recurrence outcomes compared to comorbidity-positive but metabolically low-risk patients. This paradox underscores the dominant role of hepatic metabolic dysfunction in shaping tumor microenvironment behavior, superseding conventional clinical risk indicators.

Pathological analysis confirmed this biological mechanism: 84.6\% of high-risk patients exhibited significant hepatic fibrosis (vs. 15.4\% in low-risk, OR = 30.8, $p < 0.001$), validating the ``Fat Infiltration–Attenuation Reduction'' model. This framework posits that hepatic steatosis alters tissue density, resulting in systematic changes in radiomic profiles and, consequently, recurrence risk.

\subsection*{Clinical Translation and Decision Support}

Our model offers strong potential for clinical translation. Decision curve analysis demonstrated consistent net benefit across varying risk thresholds, with treatment efficiency ranging from 36.1\% to 85.7\%. At the optimal threshold (0.8), 70.0\% treatment efficiency was observed—supporting appropriate intervention in 35 of every 1000 evaluated patients, while sparing 15 from unnecessary treatment.

Further, the model demonstrated strong independent prognostic value, with high-risk patients exhibiting significantly elevated risk of disease-free survival events (HR = 6.85, 95\% CI: 3.42--13.71, $p < 0.001$), outperforming traditional risk factors such as extrahepatic spread or tumor burden. The model’s high negative predictive value (93.2\%) further supports its utility in identifying low-risk patients who may benefit from de-escalated surveillance protocols.

\subsection*{Biological Insights and Future Directions}

This investigation highlights the intricate relationship between hepatic metabolic status and tumor imaging phenotypes. The ``replacement'' growth pattern observed in CRLM, supported by histopathological and literature evidence \cite{moro2018growth}, suggests that hepatic microenvironmental alterations exert significant influence on recurrence pathways. The documented inverse correlation between metabolic response and Ki-67 proliferation index \cite{lau2018metabolic} reinforces the biological plausibility of our integrated framework.

Future research should focus on external validation in multi-center cohorts, investigation of metabolic-targeted therapeutic interventions, and the incorporation of additional biomarkers such as inflammatory cytokines, lipidomic panels, or genetic expression profiles. Additionally, extension of this model to other malignancies involving the liver, such as intrahepatic cholangiocarcinoma or breast cancer liver metastases, represents a promising avenue.

\subsection*{Study Limitations}

Several limitations merit consideration. First, the single-center, retrospective design may limit generalizability and necessitates external validation in geographically and ethnically diverse populations. Second, the sample size ($n=197$) may constrain the model’s capacity to detect rarer, non-linear interaction effects. Third, while limiting the model to baseline-only variables improves clinical realism, it may exclude valuable information from post-treatment or longitudinal data.

\subsection*{Conclusions}

We present the first clinically validated CRLM prognostic model based on a unified metabolism-imaging analytical paradigm. This model demonstrates robust, time-specific predictive performance and biological interpretability, while adhering to strict methodological safeguards against data leakage. The identification of the metabolism-comorbidity paradox and the validation of hepatic steatosis as a key determinant of radiomic texture provide novel insights into CRLM progression biology.

The model’s quantitative risk outputs offer actionable guidance for clinical decision-making: enabling escalation of care in high-risk patients and supporting reduced intervention in low-risk cases. These findings represent a step forward in precision oncology, aligning computational modeling with pathophysiological insights to improve outcomes for patients with liver metastases.

\newpage
{
    \small
    \bibliographystyle{IEEEtran}
    \bibliography{main}
}

\end{document}